\documentclass[10pt,final,journal,twocolumn]{IEEEtran}
\ifCLASSINFOpdf

\else
\usepackage{algorithm}
\usepackage{booktabs}

\fi
\usepackage{cite}
\usepackage[cmex10]{amsmath}
\usepackage{amsthm}
\usepackage{stfloats}
\usepackage{multicol,multienum}
\usepackage{multirow}

\usepackage{amssymb}
\usepackage{url}
\usepackage{amsmath}
\usepackage{xcolor}
\usepackage[dvips]{graphicx}
\usepackage{subfigure}
\usepackage{caption}
\usepackage{amsmath}
\usepackage{amsfonts,amssymb}
\usepackage{bbm}
\usepackage[T1]{fontenc}
\usepackage{algorithm}
\usepackage{algpseudocode}
\usepackage{ulem}

\usepackage[linesnumbered, ruled]{}
\makeatletter
\renewcommand{\maketag@@@}[1]{\hbox{\m@th\normalsize\normalfont#1}}%
\makeatother

\begin{document}
\hyphenation{op-tical net-works semi-conduc-tor}
\title{Intelligent Reflecting Surface-Aided Wireless
Communication with Movable Elements}
\author{Guojie Hu, Qingqing Wu,~\textit{Senior Member}, \textit{IEEE}, Dognhui Xu, Kui Xu,~\textit{Member}, \textit{IEEE}, Jiangbo Si,~\textit{Senior Member}, \textit{IEEE}, Yunlong Cai,~\textit{Senior Member}, \textit{IEEE}, and Naofal Al-Dhahir,~\textit{Fellow}, \textit{IEEE}
\thanks{
%
Guojie Hu and Donghui Xu are with the College of Communication Engineering, Rocket Force University of Engineering, Xi'an 710025, China (email: lgdxhgj@sina.com). Qingqing Wu is with the Department
of Electronic Engineering, Shanghai Jiao Tong University, Shanghai 200240, China. Kui Xu is with the College of Communications Engineering, the Army of Engineering University, Nanjing 210007, China. Jiangbo Si is with the Integrated Service Networks Lab of Xidian University, Xi'an 710100, China. Yunlong Cai
is with the College of Information Science and Electronic Engineering,
Zhejiang University, Hangzhou 310027, China. Naofal Al-Dhahir is with the Department of Electrical and Computer Engineering, The University of Texas at Dallas, Richardson, TX 75080 USA.
}
}
\IEEEpeerreviewmaketitle
\maketitle
\begin{abstract}
Intelligent reflecting surface (IRS) has been recognized as a powerful technology for boosting communication performance. To reduce manufacturing and control costs, it is preferable to consider discrete phase shifts (DPSs) for IRS, which are set by default as uniformly distributed in the range of $[ - \pi ,\pi )$ in the literature. Such setting, however, cannot achieve a desirable performance over the general Rician fading where the channel phase concentrates in a narrow range with a higher probability. Motivated by this drawback, we in this paper design optimal non-uniform DPSs for IRS to achieve a desirable performance level. The fundamental challenge is the \textit{possible offset in phase distribution across different cascaded source-element-destination channels}, if adopting conventional IRS where the position of each element is fixed. Such phenomenon leads to different patterns of optimal non-uniform DPSs for each IRS element and thus causes huge manufacturing costs especially when the number of IRS elements is large. Driven by the recently emerging fluid antenna system (or movable antenna technology), we demonstrate that if the position of each IRS element can be flexibly adjusted, the above phase distribution offset can be surprisingly eliminated, leading to the same pattern of DPSs for each IRS element. Armed with this, we then determine the form of unified non-uniform DPSs based on a low-complexity iterative algorithm. Simulations show that our proposed design significantly improves the system performance compared to competitive benchmarks.
\end{abstract}

\begin{IEEEkeywords}
IRS, non-uniform discrete phase shifts, phase distribution offset, movable reflecting elements.
\end{IEEEkeywords}

\IEEEpeerreviewmaketitle
\vspace{-5pt}
\section{Introduction}
Intelligent reflecting surface (IRS), with the ability of reconfiguring wireless channel environments via only low-cost reflecting elements, has been recognized as one potential technology for achieving high spectrum and energy efficiency in the next generation of wireless communications [1]$-$[2].

To fully reap the passive reflecting beamforming gains, current works mostly assume the availability of continuous phase shifts (CPSs) at each IRS reflecting element [3]$-$[6]. Nevertheless, it is difficult to employ CPSs since manufacturing and controlling each reflecting element with more phase shifts leads to a higher cost, especially when the number of reflecting elements is very large [7].

Responding to this, subsequent works have considered more practical discrete phase shifts (DPSs) settings at the IRS, see [8]$-$[12]. However, all related works just adopt the uniformly distributed DPSs at each IRS element, i.e., $\left\{ { - \pi , - \pi  + \frac{{2\pi }}{M}, - \pi  + \frac{{4\pi }}{M},...,\pi  - \frac{{2\pi }}{M}} \right\}$ ($M$ is the number of available phase shifts), to quantify the channel phase. We must emphasize that this simple setting actually is not very effective for the general Rician fading. The key reason lies in that the phase distribution of Rician fading concentrates in a narrow range with a higher probability, instead of being uniformly distributed in $[ - \pi ,\pi )$ [13]$-$[14]. Therefore, some shifts in the uniformly distributed DPSs are somehow wasted to quantify channel phases with a very small probability of occurrence. As a comparison, non-uniformly distributed DPSs where more shifts gather around the most-likely channel phases are more attractive to minimize the average quantization error.

In this letter, we aim to determine the pattern of the optimal non-uniformly distributed DPSs at the IRS to maximize the average rate of a typical system consisting of a source, an IRS and a destination. One fundamental challenge lies in that there may exist the phase distribution offset phenomenon across different source-element-destination channels. Since the DPSs should be designed based on the channel phase distribution information, the above phenomenon will lead to totally different patterns of the optimal non-uniform DPSs for different IRS elements and then seriously increase manufacturing costs. Fortunately, motivated by the recently emerging fluid antenna system (FAS) [15]$-$[17] (which is later evolved into the movable antenna technology [18]$-$[24]), we reveal that if the position of each reflecting element can be flexibly adjusted to the specified point, the above mentioned phase offset can be completely eliminated, leading to two significant advantages: i) all reflecting elements will adopt the same pattern of the optimal non-uniform DPSs, which is then optimized via a low-complexity algorithm [14]; ii) even if the destination moves to a different location, the IRS does not need to update the pattern of the non-uniform DPSs but just re-adjust the positions of all reflecting elements accordingly, which further reduces the costs of replacing the IRS for serving users in different locations. It is examined by simulations that our proposed design can also achieve a pretty good rate performance compared to competitive benchmarks.

 \newcounter{mytempeqncnt}
 \vspace{-5pt}
\section{System Model}
As illustrated in Fig. 1, a single-antenna source (S) intends to communicate with a single-antenna destination, which is assisted by an IRS consisting of $N \ge 2$ movable reflection elements arranged in a linear array.\footnotemark \footnotetext{The IRS can also exploit the general planar array for reflection. However, our key conclusion is not affected by this change. Thus, we consider a simple setting in this paper.} There is no direct link between S and D due to unfavorable conditions such as propagation obstacles. The channel coefficients between S and the IRS and those between the IRS and D are denoted as ${{\bf{h}}_{sr}} \in {{\mathbb{C}}^{N \times 1}}$ and ${{\bf{h}}_{rd}} \in {{\mathbb{C}}^{1 \times N}}$, respectively. In practice, S usually represents a base station with its antenna located at a higher height, and the IRS is assumed to be attached to a surrounding building's facade. Therefore, it is reasonable to assume that the channel from S to the IRS is dominated by the line-of-sight (LoS) link. On the other hand, D generally represents a typical user located on the ground and there may exist less shadowing but non-negligible small-scale fading between the IRS and D. Thus, the IRS-D link should be characterized by a Rician fading model. Based on these, it is determined that
 \begin{figure}
\centering
\includegraphics[width=8cm]{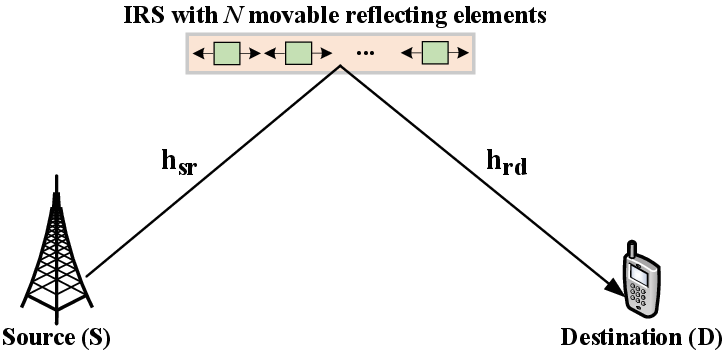}
\captionsetup{font=small}
\caption{Illustration of the system model.} \label{fig:Fig1}
\end{figure}
\begin{equation}
\begin{split}{}
{{\bf{h}}_{sr}} =& \sqrt {\frac{{{\beta _0}}}{{d_{sr}^\alpha }}} {e^{ - j\frac{{2\pi {d_{sr}}}}{\lambda }}}\\
 &\times {\left[ {{e^{ - j\frac{{2\pi {x_1}}}{\lambda }\cos {\varphi _{sr}}}},...,{e^{ - j\frac{{2\pi {x_N}}}{\lambda }\cos {\varphi _{sr}}}}} \right]^T},
\end{split}
\end{equation}
where ${\left[  \cdot  \right]^T}$ is the transpose operation, ${{\beta _0}}$ denotes the channel power at the reference distance of 1 meter, ${d_{sr}}$ denotes the distance between S and the IRS, $\alpha $ is the path loss exponent, $\lambda $ is the carrier wavelength, ${{\varphi _{sr}}}$ is the angle of arrival (AoA) to the IRS, and $\left\{ {{x_1},{x_2},...,{x_N}} \right\}$ are the positions of $N$ movable elements relative to the reference point zero, which can be flexibly adjusted in the specified region for enhancing the performance.\footnotemark \footnotetext{It should be emphasized that, similar to [18]$-$[24], the far-field condition is assumed between S and the IRS since the size of the moving region for all reflecting elements is much smaller than the signal propagation distance. Thus, the AoAs and the amplitudes of the complex coefficients for multiple channel paths do not change for different positions of the reflecting elements.} The details are shown later. In addition,
\begin{equation}
\begin{split}{}
{{\bf{h}}_{rd}} = \sqrt {\frac{{{\beta _0}}}{{d_{rd}^\alpha }}} \left( {\sqrt {\frac{K}{{K + 1}}} {\bf{h}}_{rd}^{{\rm{LoS}}} + \sqrt {\frac{1}{{K + 1}}} {\bf{h}}_{rd}^{\rm{{NLoS}}}} \right),
\end{split}
\end{equation}
where ${d_{rd}}$ is the distance between the IRS and D, $K$ is the Rician factor and ${{\bf{h}}_{rd}^{{\rm{LoS}}}}$ is the LoS component, with
\begin{equation}
{\bf{h}}_{rd}^{{\rm{LoS}}} = {e^{ - j\frac{{2\pi {d_{rd}}}}{\lambda }}}\left[ {{e^{ - j\frac{{2\pi {x_1}}}{\lambda }\cos {\varphi _{rd}}}},...,{e^{ - j\frac{{2\pi {x_N}}}{\lambda }\cos {\varphi _{rd}}}}} \right],
\end{equation}
where ${{\varphi _{rd}}}$ is the angle of departure (AoD) from the IRS to D. In addition, ${{\bf{h}}_{rd}^{{\rm{NLoS}}}}$ is the non-LoS (NLoS) component, where each element of ${{\bf{h}}_{rd}^{{\rm{NLoS}}}}$ is i.i.d. complex Gaussian distributed with zero mean and unit variance.

Based on the above analysis, the received signal at D is expressed as
\begin{equation}
{y_d} = \sqrt {{P_s}} {{\bf{h}}_{rd}}{\bf{\Theta }}{{\bf{h}}_{sr}}{x_s} + {n_d},
\end{equation}
where $P_s$ is the transmit power of S, ${x_s} \sim {\cal C}{\cal N}(0,1)$ is the transmit signal of S, ${\bf{\Theta }} = {\rm{diag}}\left\{ {{e^{j{\theta _1}}},{e^{j{\theta _2}}},...,{e^{j{\theta _N}}}} \right\}$ is the phase shift matrix, ${\theta _i}$ is the phase shift of the $i$th IRS reflecting element, and $n_d$ is the additive complex Gaussian noise with zero mean and variance of ${{\sigma ^2}}$. By denoting ${{\bf{h}}_{sr}} = {\left[ {{h_{sr,1}},...,{h_{sr,N}}} \right]^T}$, ${{\bf{h}}_{rd}} = \left[ {{h_{rd,1}},...,{h_{rd,N}}} \right]$ and ${\bf{h}}_{rd}^{{\rm{NLoS}}} = \left[ {h_{rd,1}^{{\rm{NLoS}}},...,h_{rd,N}^{{\rm{NLoS}}}} \right]$, $y_d$ in (4) is expanded as
\begin{equation}
\begin{split}{}
{y_d} =& \sqrt {{P_s}} \sum\nolimits_{i = 1}^N {{h_{rd,i}}{h_{sr,i}}{e^{j{\theta _i}}}{x_s}}  + {n_d}\\
 =& \frac{{\sqrt {{P_s}} {\beta _0}}}{{\sqrt {d_{sr}^\alpha d_{rd}^\alpha } }}\sum\nolimits_{i = 1}^N {{e^{j\left( {{\theta _i} - \frac{{2\pi {x_i}}}{\lambda }\cos {\varphi _{sr}}} \right)}}}   \\
 \times& \left( {\sqrt {\frac{K}{{K + 1}}} {e^{ - j\frac{{2\pi {x_i}}}{\lambda }\cos {\varphi _{rd}}}} + \sqrt {\frac{1}{{K + 1}}} h_{rd,i}^{{\rm{NLoS}}}} \right) + {n_d}\\
 =& \frac{{\sqrt {{P_s}} {\beta _0}}}{{\sqrt {d_{sr}^\alpha d_{rd}^\alpha } }}\sum\nolimits_{i = 1}^N {{e^{j\left( {{\theta _i} - \frac{{2\pi {x_i}}}{\lambda }(\cos {\varphi _{sr}} + \cos {\varphi _{rd}})} \right)}}}   \\
 \times& \left( {\underbrace {\sqrt {\frac{K}{{K + 1}}}  + \sqrt {\frac{1}{{K + 1}}} h_{rd,i}^{{\rm{NLoS}}}{e^{j\frac{{2\pi {x_i}}}{\lambda }\cos {\varphi _{rd}}}}}_{{h_i}}} \right){x_s} + {n_d},
\end{split}
\end{equation}
leading to the receiving signal-to-noise ratio (SNR) of D as
\begin{equation}
{\gamma _d} = \frac{{{P_s}\beta _0^2}}{{d_{sr}^\alpha d_{rd}^\alpha {\sigma ^2}}}{\left| {\sum\nolimits_{i = 1}^N {\left| {{h_i}} \right|{e^{j\left( {{\varphi _{{h_i}}} - \frac{{2\pi {x_i}}}{\lambda }(\cos {\varphi _{sr}} + \cos {\varphi _{rd}}) + {\theta _i}} \right)}}} } \right|^2},
\end{equation}
where ${{\varphi _{{h_i}}}}$ is the phase of ${{h_i}}$.

Observing (6), if CPSs are adopted, i.e., each ${{\theta _i}}$, $i = 1,...,N$, can be set to any value within the interval $[ - \pi ,\pi )$, ${\gamma _d}$ can achieve its maximum given ${\theta _i} = \frac{{2\pi {x_i}}}{\lambda }(\cos {\varphi _{sr}} + \cos {\varphi _{rd}}) - {\varphi _{{h_i}}} + c$, where $c$ is an arbitrary constant. However, considering the manufacturing and control costs, generally there are only finite DPSs available for each IRS element. Then, the key problem is how to determine the pattern of such DPSs to achieve a desirable performance? Specifically, the phase of the cascaded S-$i$th-element-D channel, i.e., ${{\varphi _{{h_i}}} - \frac{{2\pi {x_i}}}{\lambda }(\cos {\varphi _{sr}} + \cos {\varphi _{rd}})}$, is randomly changed in each transmission block because ${{\varphi _{{h_i}}}}$ is affected by the random NLoS component ${h_{rd,i}^{{\rm{NLoS}}}}$. Therefore, the pattern of the corresponding DPSs cannot be designed based on one realization of the above phase but rather its statistical information. Next, we will present a fundamental challenge of the phase distribution offset in designing the pattern of the DPSs and then provide an effective solution by flexibly adjusting the positions of all reflecting elements.

\newcommand{\tabincell}[2]{\begin{tabular}{@{}#1@{}}#2\end{tabular}}
\renewcommand\arraystretch{1.35}
\begin{table*}[t]
\caption{The Pattern of The Optimal DPSs under Different Rician Factor $K$.}
\scriptsize
\centering
 \begin{tabular}[l]{|c|c|c|c|c|c|c|c|}
 \hline
 \normalsize System Parameters&\normalsize Optimal DPSs \\
\hline
 \normalsize $K = 2,M = 4$&\normalsize $- 3.1415{\rm{ }} \ - {\rm{0}}{\rm{.5201 }} \ \ 0 \ \ {\rm{0.5201 }}$\\
 \hline
 \normalsize $K = 4,M = 4$&\normalsize $- 3.1415{\rm{ }} \ - {\rm{0}}{\rm{.3672 }} \ \ 0 \ \ {\rm{0.3672 }}$\\
 \hline
 \normalsize $K = 2,M = 8$&\normalsize $- 3.1415{\rm{ }} \ - {\rm{0}}{\rm{.8569 }} \ - {\rm{0}}{\rm{.4983 }} \ - {\rm{0}}{\rm{.2347}} \ \ 0 \ \ {\rm{0.2347 }} \ {\rm{0.4983}} \ {\rm{0.8569}}$\\
 \hline
 \normalsize $K = 4,M = 8$&\normalsize $- 3.1415{\rm{ }} \ - {\rm{0}}{\rm{.6048 }} \ - {\rm{0}}{\rm{.3517 }} \ - {\rm{0}}{\rm{.1657}} \ \ 0 \ \ {\rm{0.1657 }} \ {\rm{0.3517}} \ {\rm{0.6048}}$\\
 \hline
 \end{tabular}
{
 \label{tab:tb1}
}
\end{table*}

\section{The Phenomenon of Phase Distribution Offset and An Effective Solution}
As explained in Section II, to design the optimal pattern of DPSs, it is necessary to first figure out the distribution of the phase ${{\varphi _{{h_i}}} - \frac{{2\pi {x_i}}}{\lambda }(\cos {\varphi _{sr}} + \cos {\varphi _{rd}})}$ of the cascaded S-$i$th-element-D channel for any $i = 1,...,N$. Specifically, from (3) it is determined that ${\varphi _{{h_i}}}$ has the same distribution for any $i = 1,...,N$, because the modulus of the complex ${e^{j\frac{{2\pi {x_i}}}{\lambda }\cos {\varphi _{sr}}}}$ for any $i = 1,...,N$ always equals one and then ${h_{rd,i}^{{\rm{NLoS}}}{e^{j\frac{{2\pi {x_i}}}{\lambda }\cos {\varphi _{rd}}}}}$ is still complex Gaussian distributed with zero mean and unit variance. However, to avoid the coupling effect [18], $x_i$ is different from $x_k$ for any $i \ne k$, which implies that $\frac{{2\pi {x_i}}}{\lambda }(\cos {\varphi _{sr}} + \cos {\varphi _{rd}}) \ne \frac{{2\pi {x_k}}}{\lambda }(\cos {\varphi _{sr}} + \cos {\varphi _{rd}})$ for any $i \ne k$. In light of the above two aspects, it is determined that if $\frac{{2\pi {x_i}}}{\lambda }(\cos {\varphi _{sr}} + \cos {\varphi _{rd}}) \ne \frac{{2\pi {x_k}}}{\lambda }(\cos {\varphi _{sr}} + \cos {\varphi _{rd}}) + 2m\pi $ for any $m = 0, \pm 1, \pm 2,...$, there will exist the distribution offset phenomenon for the phases of S-$i$th-element-D and S-$k$th-element-D channels. For better understanding, we here provide an example as shown in Fig. 2, where we set $K = 10$, ${x_i} = (i - 1)\lambda /2$ and $\cos {\varphi _{sr}} + \cos {\varphi _{rd}} = 0.2$. From Fig. 2, it is clear that there exists an offset of $\frac{{2\pi \lambda }}{{2\lambda }}(\cos {\varphi _{sr}} + \cos {\varphi _{rd}}) = 0.2\pi $ for the phase distributions of two adjacent S-element-D channels.

 \begin{figure}
\centering
\includegraphics[width=7cm]{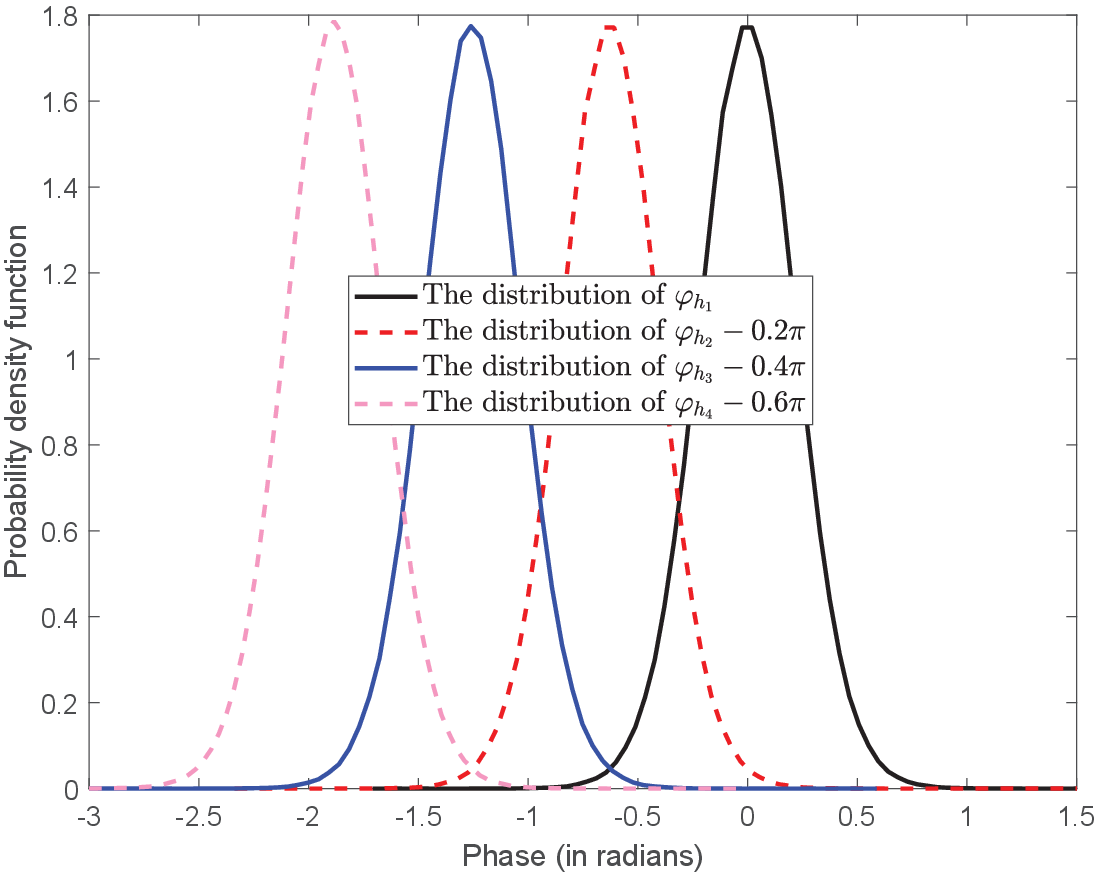}
\captionsetup{font=small}
\caption{The distribution offset phenomenon for the phases of different S-element-D channels.} \label{fig:Fig1}
\end{figure}

\subsection{Why the phase distribution offset phenomenon is harmful?}
For the cascaded S-$i$th-element-D channel, the corresponding pattern of DPSs of the $i$th reflecting element should be reasonably designed based on the phase distribution information of such cascaded channel. As a sequence, if there exists the phase distribution offset across different cascaded channels, the patterns of the optimal DPSs for different reflecting elements will be totally different. For instance, focusing on Fig. 2, if the optimal pattern of DPSs for the first element of the IRS is $\left\{ { - \pi , - \frac{\pi }{9},0,\frac{\pi }{9}} \right\}$ given the number of shifters is 4, the optimal pattern of DPSs for the second element would become $\left\{ { - \pi , - \frac{\pi }{9},0,\frac{\pi }{9}} \right\} - 0.2\pi $, and for the third element becomes $\left\{ { - \pi , - \frac{\pi }{9},0,\frac{\pi }{9}} \right\} - 0.4\pi $ and so on. More importantly, if the AoA and the AoD are changed to enable $\cos {\varphi _{sr}} + \cos {\varphi _{rd}} = 0.3$, the offset for the phase distributions of two adjacent S-element-D channels would become $0.3\pi $. This causes the optimal patterns of DPSs for the second and third elements to become $\left\{ { - \pi , - \frac{\pi }{9},0,\frac{\pi }{9}} \right\} - 0.3\pi $ and $\left\{ { - \pi , - \frac{\pi }{9},0,\frac{\pi }{9}} \right\} - 0.6\pi $, respectively.

Based on the above analysis, even when the AoA (${\varphi _{sr}}$) and the AoD (${\varphi _{rd}}$) are fixed, the patterns of the optimal DPSs for different reflecting elements are different, which causes huge manufacturing costs as there is no unified production line to produce changeable DPSs. More importantly, once the AoA and the AoD vary, e.g., the served user is in a different location, the optimal pattern of DPSs for each reflecting element will change inevitably. This further poses a huge challenge for an IRS with fixed patterns of DPSs to optimally serve the user in any position.

\subsection{Our proposed effective solution}
Inspired by the recently emerging movable antenna technology, we now assume that the positions of the $N$ reflecting elements can be flexibly adjusted. Actually, as pointed out in [15]$-$[24], the positions of all reflecting elements can be controlled by step motors or servos. By receiving the control signal from the central processing unit, the motors can perform the corresponding steps to move each element to a target position with a desired accuracy in tenth of the wavelength.

Without loss of optimality, $x_1$ is set as zero, and ${x_1} < {x_2} < {x_3} < ... < {x_N}$. Via the above analysis, to completely eliminate the phase distribution offset, the condition $\frac{{2\pi {x_i}}}{\lambda }\left| {\cos {\varphi _{sr}} + \cos {\varphi _{rd}}} \right| = 2(i - 1)\pi $, for any $i = 1,...,N$, should be satisfied, leading to
\begin{equation}
{x_i} = \frac{{(i - 1)\lambda }}{{\left| {\cos {\varphi _{sr}} + \cos {\varphi _{rd}}} \right|}}, i = 1,...,N.
\end{equation}

Given the AoA and the AoD, the $N$ reflecting elements just need to move to their respective locations based on (7) and then stay unchanged permanently. Then, the phase distribution characteristics of all $N$ cascaded S-element-D channels will be exactly the same, and the corresponding probability density function (PDF) can be determined as the black curve in Fig. 2. The benefit of this design is obvious: the pattern of the optimal DPSs for all reflecting elements will be the same and then the production cost and time can be reduced significantly. More importantly, even when the AoA and the AoD change, the IRS never needs to update the patterns of DPSs for all reflecting elements but just change the positions of all reflecting elements based on (7). This indicates that the proposed design is robust to changes in the user position.

\section{Optimal Non-Uniform DPSs and Average Rate}
\subsection{Optimal Non-Uniform DPSs}
After eliminating the phase distribution offset, the receiving SNR of D becomes
\begin{equation}
{\gamma _d} = \frac{{{P_s}\beta _0^2}}{{d_{sr}^\alpha d_{rd}^\alpha {\sigma ^2}}}{\left| {\sum\nolimits_{i = 1}^N {\left| {{h_i}} \right|{e^{j\left( {{\varphi _{{h_i}}} + {\theta _i}} \right)}}} } \right|^2}.
\end{equation}

Note that ${\gamma _d}$ in equation (8) has the same form as ${\gamma _d}$ in equation (4) of our recent work [14]. Hence, the low-complexity Algorithm 1 developed in [14] can be exploited to find the optimal DPSs. The details are omitted here as these are not the main focus of this paper. As shown in Table I, we just present the computed optimal DPSs under different Rician-factor $K$, where $M$ denotes the number of available phase shifters at each IRS element. Table I will be exploited for subsequent simulations. From Table I, it is clear to observe that the optimal DPSs are not uniformly distributed. This is because for the general Rician fading, as presented in Fig. 2, the channel phase ${{\varphi _{{h_i}}}}$ concentrates in a narrow range with a higher probability instead of being uniformly distributed in $[ - \pi ,\pi )$, which indicates that more shifters should gather around the most-likely phases to minimize the quantization error. We will show the performance gain of the non-uniformly distributed DPSs obtained in this paper compared to the widely used uniformly distributed DPSs.

\subsection{Average Rate}
Based on the obtained non-uniform DPSs, in each transmission block, the selection of the specific DPS (denoted as $\theta _i^*$) at the $i$th IRS reflecting element is based on the rule of $\theta _i^* = \arg \min \left| {{\varphi _{{h_i}}} + {\theta _i}} \right|$ [8], [13]$-$[14]. By denoting the residual phase error as ${\phi _i} = {\varphi _{{h_i}}} + \theta _i^*$, the average rate (denoted as $\overline C $) and its upper bound are
\begin{equation}
\begin{split}{}
\overline C  =& {\mathbb{E}}\left[ {{{\log }_2}\left( {1 + {\gamma _d}} \right)} \right]\\
 =& {\mathbb{E}}\left[ {{{\log }_2}\left( {1 + \frac{{{P_s}\beta _0^2}}{{d_{sr}^\alpha d_{rd}^\alpha {\sigma ^2}}}{{\left| {\sum\nolimits_{i = 1}^N {\left| {{h_i}} \right|{e^{j{\phi _i}}}} } \right|}^2}} \right)} \right]\\
\mathop  \le \limits^{(a)} &{\log _2}\left( {1 + \frac{{{P_s}\beta _0^2}}{{d_{sr}^\alpha d_{rd}^\alpha {\sigma ^2}}}{\mathbb{E}}\left[ {{{\left| {\sum\nolimits_{i = 1}^N {\left| {{h_i}} \right|{e^{j{\phi _i}}}} } \right|}^2}} \right]} \right),
\end{split}
\end{equation}
where $\mathop  \le \limits^{(a)} $ is due to the Jensen's inequality and
\begin{equation}
\begin{split}{}
&{\mathbb{E}}\left[ {{{\left| {\sum\nolimits_{i = 1}^N {\left| {{h_i}} \right|{e^{j{\phi _i}}}} } \right|}^2}} \right] = {\left( {N{\Omega _{1,0.5}}{\Omega _{2,1}}} \right)^2}\\
 -& N\left( {{\Omega _{1,1}}{\Omega _{2,2}} - \Omega _{1,0.5}^2\Omega _{2,1}^2} \right) - N{\Omega _{1,1}}{\Omega _{3,2}},
\end{split}
\end{equation}
where the parameters ${{\Omega _{1,0.5}}}$, ${{\Omega _{1,1}}}$, ${{\Omega _{2,1}}}$, ${{\Omega _{2,2}}}$ and ${\Omega _{3,2}}$ are presented in Appendix A of [14] and thus are not repeated here.

\section{Simulation Results}
In this section, we present numerical results to evaluate the proposed design. The transmit power of S is 30 dBm. The distances between S and the IRS and between the IRS and D are ${d_{sr}} = 30$ m and ${d_{rd}} = 10$ m, the path loss exponent is $\alpha  = 3$, the channel power at the reference distance of 1 meter is ${\beta _0} = {10^{ - 3}}$, the noise power is ${\sigma ^2} =  - 110$ dBm, and the Rician factor is $K = 4$.

For comprehensive comparisons, we further consider the following competitive schemes:
\begin{itemize}
\item CPSs: In this scheme, the phase shifts at all reflecting elements are optimally set to align the phases of all $N$ cascaded channels. The resulting SNR of D is ${\gamma _d} = \frac{{{P_s}\beta _0^2}}{{d_{sr}^\alpha d_{rd}^\alpha {\sigma ^2}}}{\left| {\sum\nolimits_{i = 1}^N {\left| {{h_i}} \right|} } \right|^2}$.
\item Movable reflecting elements with uniformly distributed DPSs (ME-UDPSs): In this scheme, the IRS can also flexibly adjust the positions of all reflecting elements to fully eliminate the phase distribution offset. However, each reflecting element just adopts the uniformly distributed DPSs, i.e., $\left\{ { - \pi , - \pi  + \frac{{2\pi }}{M}, - \pi  + \frac{{4\pi }}{M},...,\pi  - \frac{{2\pi }}{M}} \right\}$, where $M$ is the number of available phase shifters.
\item Conventional IRS with uniformly distributed DPSs (C-UDPSs): In this scheme, all reflecting elements have fixed positions with ${x_i} = \frac{{(i - 1)\lambda }}{4}$, $i = 1,...,N$. In addition, the uniformly distributed DPSs are adopted.
\end{itemize}

 \begin{figure}
\centering
\includegraphics[width=7.8cm]{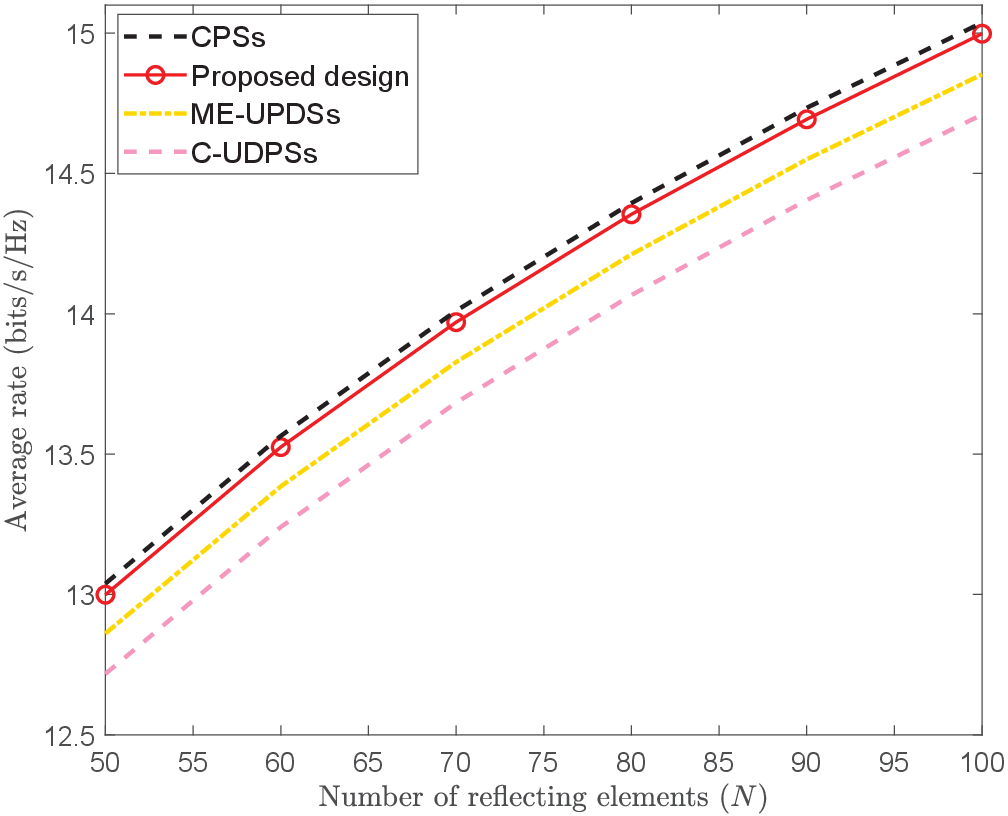}
\captionsetup{font=small}
\caption{Average rate of four schemes w.r.t. the number of reflecting elements ($N$), where $M = 4$.} \label{fig:Fig1}
\end{figure}

 \begin{figure}
\centering
\includegraphics[width=7.8cm]{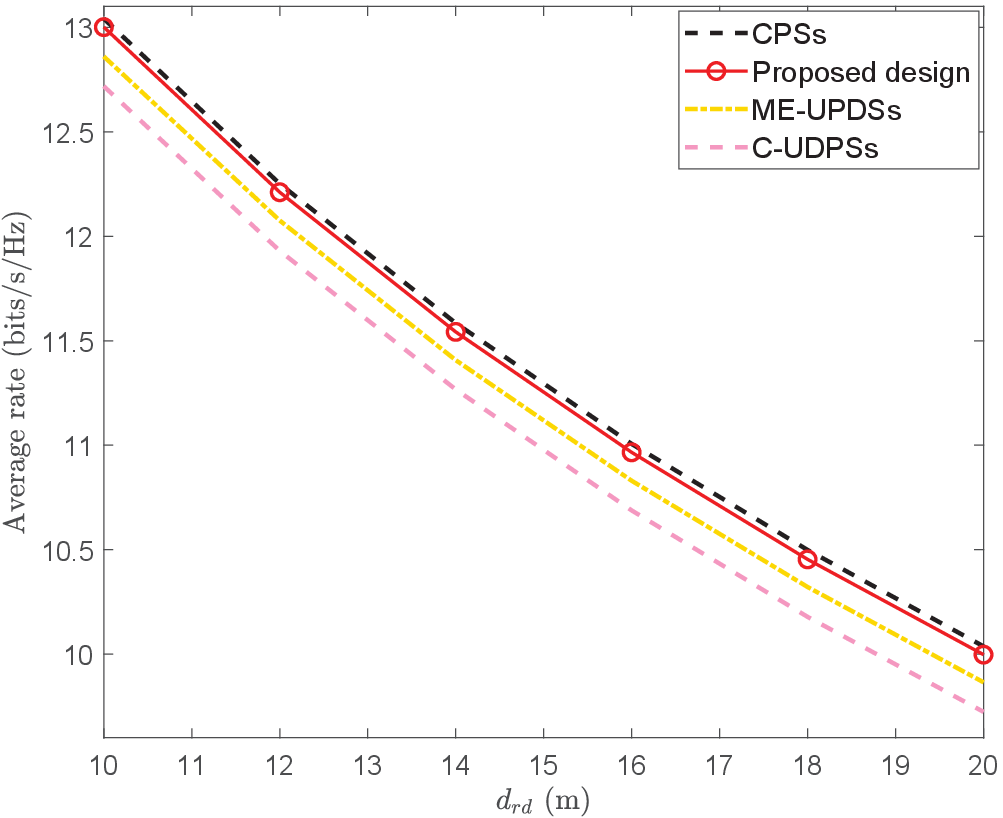}
\captionsetup{font=small}
\caption{Achievable rate of four schemes w.r.t. the distance between the IRS and D ($d_{rd}$), where $N = 50$ and $M = 4$.} \label{fig:Fig1}
\end{figure}


We aim to examine the average rate achieved by CPSs, our proposed design, ME-UDPSs and C-UDPSs. The results are obtained by averaging $10^5$ random realizations of $\left\{ {{h_i}} \right\}_{i = 1}^N$. For ME-UDPSs and C-UDPSs, in each simulation, the selection of the DPS at the $i$th reflecting element is also based on the rule of ${\min _{{\theta _i}}}\left| {{\varphi _{{h_i}}} + {\theta _i}} \right|$. Further, the performance of C-UPDSs is the average result obtained by setting $\cos {\varphi _{sr}} + \cos {\varphi _{rd}} = 0.1:0.1:1.5$.

Fig. 3 presents the average rate achieved by the four schemes versus $N$, from which it is observed: i) as $N$ increases, the reflecting beamforming gain increases. Therefore, the average rate of all schemes increases with respect to (w.r.t.) $N$; ii) for CPSs, since the continuous shifters are always available at each IRS element to align $N$ cascaded channel phases, this scheme can achieve the performance upper bound; iii) for ME-UDPSs, although the phase distribution offset is eliminated, the conventional uniformly distributed DPSs are undesirable for quantifying the non-uniformly distributed cascaded channel phase, i.e., they cause a higher quantization error, leading to the inferior performance compared to our proposed design; iv) for C-UDPSs, the phase distribution offset phenomenon is existed but at each IRS element, the corresponding DPSs are simply set as uniformly distributed, which does not respond well to the numerous different phase distributions among different cascaded channels; v) as a comparison, since the phase distribution offset is eliminated and the non-uniform DPSs are well selected, our proposed design achieves a pretty good performance.

Fig. 4 further shows the average rate w.r.t. $d_{rd}$, from which it is observed that: i) as the distance between the IRS and D increases, the average gain of the cascaded S-IRS-D channel reduces. Therefore, obviously the average rate decreases w.r.t. $d_{rd}$; ii) even when $M = 4$, there is almost no rate gap between CPSs and our proposed design, implying that a few number of phase shifters is enough to achieve a close-to-optimal performance.
%

\section{Conclusion}
This letter first reveals an inherent phenomenon of the phase distribution offset among different S-element-D channels, if adopting the conventional IRS where the positions of all reflecting elements are fixed. Then, inspired by the movable antenna technology, the positions of all reflecting elements are optimized to completely eliminate the phase offset phenomenon. Afterwards, the unified non-uniform DPSs for all reflecting elements are designed. Numerical results show that our proposed design outperforms competitive benchmarks.

\end{document}